\documentclass[12pt]{article}
\usepackage{epsf}

\font\twlmsy=msbm10 at 12pt
\font\sevenmsy=msbm8
\font\fivemsy=msbm6
\newfam\Bbbfam
\textfont\Bbbfam=\twlmsy
\scriptfont\Bbbfam=\sevenmsy
\scriptscriptfont\Bbbfam=\fivemsy

\newcommand{\rf}[1]{(\ref{#1})}
\newcommand{\oh}{\frac{1}{2}}

\newcommand{\beq}{\begin{equation}}
\newcommand{\eeq}{\end{equation}}
\newcommand{\bea}{\begin{eqnarray}}
\newcommand{\eea}{\end{eqnarray}}
\newcommand{\beas}{\begin{eqnarray*}}
\newcommand{\eeas}{\end{eqnarray*}}
\newcommand{\beqs}{\begin{displaymath}}
\newcommand{\eeqs}{\end{displaymath}}

\newcommand{\bl}{\bar{l}}

\newcommand{\ben}{\begin{equation}}
\newcommand{\een}{\end{equation}}

\newcommand{\bdm}{\begin{displaymath}}
\newcommand{\edm}{\end{displaymath}}

\newcommand{\cS}{{\cal S}}

\begin{document}
\topmargin 0pt
\oddsidemargin 5mm
\headheight 0pt
\topskip 0mm

\addtolength{\baselineskip}{0.4\baselineskip}

\pagestyle{empty}

\vspace{1cm}

\hfill RH-19-97

\vspace{2cm}

\begin{center}

{\Large \bf On the Width of Handles in Two-dimensional Quantum Gravity}



\vspace{1.2 truecm}

\vspace{1.2 truecm}

{\em Thordur Jonsson\footnote{e-mail: thjons@raunvis.hi.is} }

\bigskip

Raunvisindastofnun Haskolans, University of Iceland \\
Dunhaga 3, 107 Reykjavik \\
Iceland

\vspace{1.2 truecm}

\end{center}

\noindent
{\bf Abstract.}  We discuss the 
 average length $\bar{l}$ of the shortest non-contractible loop on
surfaces in
the two-dimensional pure quantum gravity ensemble.  The value of $\gamma
_{str}$ and the explicit form of the continuum loop functions indicate
that $\bar{l}$ diverges at the critical point.  Scaling arguments
suggest that the critical exponent of $\bar{l}$ is $\oh$.  We show that
this value of the critical exponent is also 
obtained for branched polymers with loops 
where the calculation is straightforward.

\vfill

\newpage
\pagestyle{plain}

In spite of the fact that one can calculate the continuum loop functions of
two-dimensional quantum gravity more or less explicitly \cite{book} 
we do not have a
very good understanding of what the generic surfaces contributing to the
loop functions really look like.  
In some sense these surfaces are ``thick'' since they are
not like trees or branched polymers and the ``minimal baby universe''
(minbu) picture of \cite{jain} has provided a valuable insight and some
quantitative understanding.

In this note we discuss the beahavior of a very simple quantity
which can be regarded as a measure of how far surfaces are
from being branched polymers.  This is the length $l$ of the shortest
non-contractible loop so we have in mind surfaces of genus $g>0$.  
We consider randomly triangulated surfaces and loops which consist
entirely of links on the surface.  Each link is defined to have length 1
so the length of a loop is the number of links it contains assuming
that each link is traversed only once.  It is clear that every homotopy
class of loops on a triangulated surface contains a link loop of
shortest length.  Let $\cS$ denote the class of surfaces under
consideration, let $\cS _A$ be the class of 
surfaces of area (=number of triangles) $A$ and let $|S|$ denote the
area of a surface $S$.
The length of the shortest noncontractible loop on $S$ will be denoted
$l(S)$.  
We denote the average value
of $l$ in the canonical ensemble by $\bl _A$ and in the grand canonical
ensemble by $\bl (\mu )$.  These averages are defined as
\beq
\bl _A=N(A)^{-1}\sum _{S\in\cS _A}l(S)
\eeq
and
\beq
\bl (\mu )=Z(\mu )^{-1}\sum _{S\in \cS} l (S)\,e^{-\mu |S|}
\eeq
where $N(A)=\# \cS _A$ and 
\beq
Z(\mu )=\sum _{S\in \cS}e^{-\mu |S|}
\eeq
is the partition function in the ensemble under study.  

There are several reasons why it is natural to expect the average $\bl
(\mu)$ to be large as $\mu \to\mu _0$.  
First recall that the number $N_g(A)$ of closed 
surfaces of genus $g$ made up of $A$ triangles, one of which is marked,
grows for large $A$ as
\beq
N_g(A)\sim A^{\gamma (g)-2}e^{\mu _0A}
\eeq
where $\gamma (g)=-\oh +{5\over 2}g$ and $\mu _0$ is independent of $g$.
 If the typical handle on a genus $g$ surface were thin so it could be
created by identifying two small regions on a surface of genus $g-1$
then we would expect $\gamma (g)$ to grow like $2g$ since
the entropy associated with choosing a small group of triangles is $A$.
For convenience
we shall assume from now on 
that the surfaces we discuss have a marked triangle but
this is not essential.

Additional evidence for the divergence of $\bl (\mu )$ comes from the
explicit calculation of the continuum loop functions of 2d quantum
gravity \cite{ambjornJM,book}.  If the handles were thin compared to the
diverging lengthscale in the continuum limit we would expect the loop
functions in different genera to be related by simple overall entropic
factors but this is not the case.

Perhaps the strongest argument for the macroscopic nature of handles in
2d quantum gravity is obtained by adopting the minbu picture of
\cite{jain} to estimate the width of handles.  The minbu picture is in
good accord with all numerical and analytical facts about quantum
gravity.  The estimate we are interested in is in fact implicit in
\cite{jain}.  Let us consider surfaces of genus 1 and area $A$.  
The number of these is given asymptotically by
\beq
N_1(A)\sim e^{\mu _0A}.
\eeq
Let $N_1(A;l)$ denote the number of genus 0 surfaces of area $A$ with
two boundary components of length $l$ with the property that the
boundaries cannot be deformed to
shorter boundaries on the surface.  It follows that if we identify the
two boundary components link by link we create a closed genus 1 surface
with a noncontractible loop of length $l$ which is not homotopic to a
shorter loop.  The authors of \cite{jain} argue that
\beq
N_1(A;l)\sim N_0(A)A^2l^{-1}\label{6}
\eeq
for $l\leq \sqrt{A}$ while $N_1(A;l)$ is zero for $l>\sqrt{A}$.
We can generically identify the two
boundary loops of surfaces contributing to $N_1(A,l,l)$ in $l$ different
ways.  Any surface of genus 1 (with a marked triangle) 
can be constructed uniquely in this way for some
$l$.   
It follows that
\bea
N_1(A)&\sim &\sum _{l=2}^{\sqrt{A}} lN_1(A;l)\nonumber\\
     &\sim &e^{\mu _0A}
\eea
so the picture is consistent.  
Within the approximation embodied in \rf{6} it follows
that the probability that the
shortest noncontractible loop on a torus of area $A$ has
length $l$ is given by
\beq
p_A(l)=\left\{\begin{array}{ll}A^{-1/2},~~&2\leq l\leq \sqrt{A}\\
0,~~&\mbox{otherwise.}\end{array}\right.\label{pf}
\eeq
We see that all kinematically possible lengths are equally probable.
It follows that the canonical average of $l$ is given by
\beq
\bl _A\sim\sum _{l=2}^{\sqrt{A}}{l\over\sqrt{A}}\sim\sqrt{A}.
\eeq
The corresponding grand canonical average is given by
\bea
\bl (\mu ) &=& Z(\mu )^{-1}\sum _{A,l}N_1(A)p_A(l)l\, e^{-\mu A}\nonumber\\
           &\sim&(\mu -\mu _0)^{-\oh}.\label{10}
\eea
Naively one might have expected $\bl (\mu )\sim (\mu -\mu _0 )^{-1/4}$
since the divergent correlation length defined by the two point function
of pure gravity has critical exponent $1/4$ \cite{ambjornwatabiki}.  We
will now show that there is in fact no conflict between this result and \rf{10}.

The homotopy group of a torus is generated by the equivalence classes of
two elementary loops $l_a$ and $l_b$ with the single relation 
$l_al_bl_a^{-1}l_b^{-1}=1$.  Let us adopt the convention that $l_a$ is
the shortest noncontractible loop on the torus under consideration and
we can take $l_b$ to be the shortest noncontractible loop which is not
homotopic to any power of $l_a$.  In a surface which is not thin one
would expect the lengths of $l_a$ and $l_b$ (denoted $|l_a|$ and
$|l_b|$) to be comparable while in a
branched polymer like surface $|l_a|$ is of order 1 while $|l_b|$ 
might be large. 
Let us denote the grand canonical expectation value of $|l_b|$ by 
$\bl '(\mu )$. We will argue that $\bl '(\mu )$ diverges with an exponent
$\oh$ as $\mu\to\mu _0$ so there is a symmetry between the divergence of
the lengths of $a$-loops and
$b$-loops in pure quantum gravity.  In the branched polymer phase $\bl$ is
bounded while $\bl '$ diverges as we shall see later.

We now use the pure gravity two point function $G_\mu (r)$ to estimate
$\bl '$.  The function $G_\mu (r)$ is defined as the partition function
for all surfaces with two marked triangles separated by a distance $r$. 
The asymptotic behaviour of $G_\mu (r)$ is given by
\beq
G_\mu (r)\sim r^{-3}~~~~\mbox{for}~1\ll r\ll\xi (\mu )
\eeq
and
\beq
G_\mu (r)\sim e^{-r/\xi}~~~~\mbox{for}~\xi (\mu )\ll r
\eeq
where $\xi (\mu )\sim (\mu -\mu _0)^{-1/4}$ as $\mu\to\mu_0$
\cite{ambjornwatabiki}. 
Identifying the boundaries of 
two triangles a distance $r$ apart on a planar surface
creates an elementary  noncontractible loop of length $r$.  For large
$r$ this loop must be $l_b$ so we expect
\bea
\bl '(\mu)&\sim &\sum _r r G_\mu (r)\nonumber\\
          &\sim& \xi ^{-2}(\mu )\nonumber\\
          &\sim& (\mu -\mu _0)^{-\oh}.\label{15}
\eea
We find this a suggestive argument for $\bl (\mu)\sim\bl '(\mu )$ in the
pure gravity phase while the collapse to branched polymers for central
charge $c>1$ is accompanied by a breakdown of the symmetry between
$a$-loops and $b$-loops.

Let us now turn to the calculation of the expectation value of the
length of the shortest loop
in a pure branched polymer ensemble.  We consider planar polymers (i.e.
polymers that are embedded in a plane) of
genus 1 which correspond to graphs with one loop.  On a branched polymer
any loop is noncontractible.   The
``transverse'' degree of freedom associated with the $a$-loops has been
eliminated when we consider branched polymers which are made up of
elementary links rather than narrow tubes so there is only one loop to
study.   

\begin{figure}[h]
\centerline{
{\epsfxsize=5cm \epsfbox{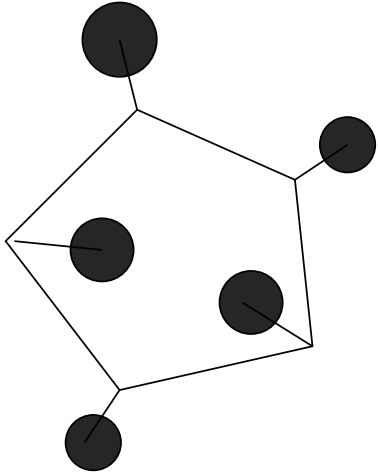}}
}
\caption{A branched polymer with a single loop of length 5.  The blobs
stand for the rooted polymer partition function and each of these
polymers sits either inside or outside the loop.}
\end{figure}

We let $Z(\beta )$ denote the 
partition function for rooted polymers with no loops.  Here $\beta$ is
the coupling constant associated with the Boltzmann weight for polymers
and we take the action to be the total number of links in the polymer. 
For simplicity let us assume that the
polymers have vertices either of order 1 or 3.  This assumption does not
affect the values of any generic critical exponents.  Let $Z_l(\beta )$ 
denote the partition function for
polymers with a single noncontractible loop of length $l$.  Then 
\beq
Z_l(\beta )=e^{-\beta l}Z(\beta )^l2^l
\eeq
since any polymer contributing to $Z_l$ can be constructed by gluing
rooted polymers to the vertices of a closed loop of length $l$, 
see Fig. 1.   The factor of $2^l$ is due to the fact that each rooted
polymer can sit inside or outside the loop.  It follows that
the average length of the loop is given by
\beq
\bl (\beta )={\sum_{l=3}^\infty lZ_l\over \sum_{l=3}^\infty Z_l}.
\eeq
Mimicking the argument used to calculate the two point function for
branched polymers \cite{bp} it is straightforward to show that
\beq
Z_l(\beta )\sim e^{-m(\beta ) l}
\eeq
as $\beta$ approaches the critical value $\beta _0$ and $m(\beta )\sim
(\beta -\beta _0)^{\oh}$ is the mass of the branched polymer model.  
It follows that
that
\beq
\bl (\beta )\sim (\beta -\beta _0)^{-\oh}.
\eeq
Note that this average is in fact analogous to the average $\bl '$
discussed for random surfaces since the transverse degree is absent for
branched polymers.

Above we have presented evidence that generic surfaces contributing to
the scaling limit of pure gravity ($c=0$) are thick in the sense that
the expectation value of the 
length of the shortest noncontractible loop diverges as the critical
point is approached.  We calculated an analogous quantity for branched
polymers and suggested that the collapse of random surfaces to a
branched polymer phase is manifested by a symmetry breaking between the
length distribution of loops in different homotopy classes.

\bigskip
\noindent
{\bf Acknowledgement.} The author is indebted to 
J. Ambj\o rn and B. Durhuus for discussions and to 
the CERN theory division for hospitality.


\begin{thebibliography}{99}
\bibitem{book}J. Ambj\o rn, B. Durhuus and T. Jonsson, Quantum geometry
- A statistical field theory approach.  Cambridge monographs on
mathematical pbysics, Cambridge University Press, Cambridge (1997).
\bibitem{jain}S. Jain and S. Mathur, Phys. Lett. B286 (1992) 239. 
\bibitem{ambjornJM}J. Ambj\o rn, J. Jurkiewicz and Y. M. Makeenko, Phys.
Lett. B251 (1990) 517.
\bibitem{ambjornwatabiki}J. Ambj\o rn and Y. Watabiki, Nucl. Phys. B445
(1995) 129.
\bibitem{bp}J. Ambj\o rn, B. Durhuus and T. Jonsson, Phys. Lett. B244
(1990) 403.

\end{thebibliography}
\end{document}